\documentclass[11pt]{article}
\usepackage{moriond,epsfig}

\def\Journal#1#2#3#4{{#1} {\bf #2}, #3 (#4)}

\def\PLB{{\em Phys. Lett.}  B}

\def\be{\begin{equation}}
\def\ee{\end{equation}}
\def\bea{\begin{eqnarray}}
\def\eea{\end{eqnarray}}

\begin{document}
\vspace*{4cm}
\title{GAUGE-HIGGS UNIFICATION IN SIX DIMENSIONS}

\author{ANDREA WULZER}

\address{ISAS-SISSA and INFN, Via Beirut 2-4, I-34013 Trieste, Italy}

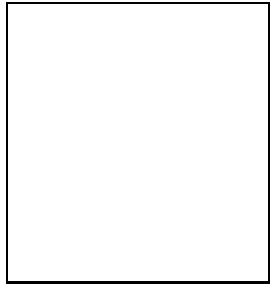
\maketitle\abstracts{Motivated by the pressing problem of the ``little hierarchies'', 
the possibility of realizing a consistent model of Electroweak Symmetry Breaking 
via the so-called ``Gauge-Higgs Unification Mechanism''
 is discussed.
We identify a class of 6 dimensional $SU(3)$ gauge  models in which a single Higgs doublet originates 
from the internal components of the gauge fields. The Higgs mass is beautifully predicted 
at the tree-level to be twice the $W$-boson mass.
At the quantum level, a 1-loop quadratically divergent localized operator is generated and it can 
contribute to the Higgs mass, reintroducing then the little hierarchy problem. 
We show that, in some particular case, the presence of this operator does not destabilize the Electroweak 
Symmetry Breaking Scale and we obtain a framework in which realistic models could be formulated.}

\section{Introduction}
\subsection{The ``Little Hierarchy Problem''}

The Standard Model of Electroweak interactions (SM) suffers from an Hierarchy Problem in the Higgs sector.
Namely, a quadratic divergence is present in the 1-loop correction to the Higgs mass-term and, 
in an effective field theory approach to the SM, the natural value of this mass is then not far from the 
cut-off itself. This requires \emph{new physics} to arise at the $TeV$ scale so as to stabilize the 
Electroweak Symmetry Breaking (EWSB) scale.

Supersymmetry (SUSY) is the standard solution to this problem. In SUSY models, the new physics 
which makes the EWSB scale stable consists on superpartners of the standard fermions, with 
supersymmetric interactions and masses lying, roughly, at the $TeV$ scale.

The problem with this picture comes from the phenomenological success of the SM in experiments 
with increasing precision and/or energy which are pushing up lower bounds on the scale of new physics 
while restricting the Higgs mass to be in the range $115-200\,GeV$. It is becoming then more and more 
difficult to find natural values for the unknown parameters of the model which are compatible 
with experimental bounds. This defines the so-called ``Little Hierarchy Problem'' \cite{bar-str} 
which affects SUSY models as well as many other extensions of the SM.

The Little Hierarchy Problem motivates the search for alternative mechanisms leading to a stable 
EWSB scale.

\subsection{Gauge-Higgs Unification}

Extra Dimensions offer many tools to address the little hierarchy problem \cite{pom-csaki}.
We will focus on the so-called ``Gauge-Higgs Unification '' (GHU) mechanism which has been proposed 
long ago and has recently received renewed interest, see \textit{e.g.} \cite{noi} and references therein.

Let us consider non-supersymmetric, $D$-dimensional gauge theories compactified to $4$ dimensions.
In general, the $D$-dimensional gauge connection will decompose in $4D$ notation as
\begin{equation}
\begin{centering}
A_M\;\rightarrow\;
\left\{
\begin{array}{c} A_\mu\; =\{\textrm{massless 4D vector bosons + K.K. towers}
\} \\ 
A_{i}\;=\{\textrm{massless 4D scalars + K.K. towers}\}
\end{array}
\right.\,,
\end{centering}
\nonumber
\end{equation}
 some scalar will then be automatically present in the $4D$ theory.

The idea, then, is to find a model in which the standard $SU(2)\times U(1)$ gauge bosons and the 
Higgs scalar field both arise from the zero modes of the extra-dimensional gauge connection. 
This is why, of course, we talk about ``Unification''. 

What is important here is that the UV dynamics of the Higgs field, which is related to the 
short-distance behavior of propagators, will be strongly constrained by the extra-dimensional 
gauge invariance. A dynamically generated Higgs mass-term, for instance,would be \emph{finite}
if the theory is compactified on a smooth manifold. The short-distance physics is indeed 
insensitive to the compactness of the extra space and on a flat $D$-dimensional Minkowski space,
which a smooth manifold locally resembles,  
gauge invariance would forbid this mass-term. At the technical level, the finiteness is ensured 
by the fact that the Higgs mass can only arise from \emph{non-local} operators in the (smooth) 
extra dimensions.

In the case of orbifold compactification, dangerous localized operators can arise at the fixed points 
\cite{viq} 
as the orbifold is not smooth at that points. Localized operators contributing to the Higgs 
mass-term will give rise to a divergence in the EWSB scale, then reintroducing an hierarchy problem.
However, we need orbifolds to obtain a $4D$-chiral spectrum and Higgs not in the adjoint of the 
Electroweak gauge group. The stabilization of the EWSB scale is then a non-trivial issue to be checked 
in any explicit model.

The case of $5$-dimensional ($5D$) models is very interesting with respect to the problem of stabilization. 
Indeed, in $5D$ the gauge symmetry forbids the presence of localized operators contributing to the 
Higgs potential \cite{viq}. For this reason, the possibility of building realistic $5D$ GHU models has been 
studied in detail.
The SM fermions are introduced in these models as boundary fields localized at the fixed points 
and a mechanism for 
generating Yukawa couplings via non-local operators has been found which can be easily 
generalized to any dimension. An extra bonus is a natural exponential hierarchy in the fermion masses.
In such models, the extra-dimensional gauge group is $SU(3)$, which is known to give too big a 
weak mixing angle. This problem has been solved by adding an extra $U(1)$ factor to the 
gauge group and making the corresponding gauge boson massive via a Green-Schwarz anomaly cancellation 
mechanism \cite{SSS}. Also this mechanism is generalizable to any dimension.
At the moment, semi-realistic $5D$ GHU models can be formulated. The basic problem is that the predicted 
Higgs mass is generally too small, smaller then the $W$-boson mass. This is due to the fact that in 
$5D$ the full Higgs potential is radiatively generated.

The 6D models we will explore  \cite{noi} can solve the problem of the Higgs mass. A quartic coupling 
for the Higgs is indeed present already at the tree-level \cite{AH}.
On the other hand, we will face problems with the 
stabilization of the EWSB scale as localized operators contributing to the Higgs mass can now arise
\cite{CGM} \cite{viq} .

\section{Gauge-Higgs Unification in $6$ Dimensions}

We will now discuss the issue of the stabilization of the EWSB scale and the size of the Higgs mass 
 in $6D$ models. 
To this end, we will study a specific class of models which 
capture the essential features of the problem. You will see that they are not realistic
as the weak mixing angle turns out to be too big and they do not contain the SM fermions at all. 
However, it would be very easy to adjust the weak angle and to introduce SM fermions and masses
 through the mechanisms we mentioned in the previous section. 
The required modifications would not change the results we will derive.

\subsection{Set-up}

Consider an $SU(3)$ gauge theory compactified on the $T^2/{\bf Z_N}$ orbifold, for the allowed values 
of $N=2,3,4,6\,$. The ${\bf Z_N}$ group acts as a $2\pi/N$ rotation ($z\rightarrow \tau z;\; 
\tau=e^{\frac{2\pi i}N}$) on the complex coordinate $z$ of $T^2$. The gauge connection 
$A_M\equiv A_{M,A}t^A\; (M=\mu,\; z,\; \bar z)$ is periodic along the cycles of $T^2$, but subjected 
to the ``twisted'' boundary conditions under the orbifold action
\begin{equation}
\label{bc}
A_M(\tau z)=R_{M}^{\; N}PA_N(z)P^{\dag}\,,
\end{equation}
where $R$ is the representation on vectors of $2\pi/N$ rotations in the $z$ plane
\begin{equation}
R_{\mu}^{\; \nu}=\delta_{\mu}^\nu\,,\;\;\;\;\; R_{z}^{\; z}=(R_{\bar z}^{\; \bar z})^*=\bar\tau\,,
\end{equation}
and $P$ is a gauge twist matrix we choose to be
\begin{equation}
P=\left(\begin{array}{ccc} \tau & 0   & 0\\
                       0  &\tau & 0\\
                       0  & 0   & 1
\end{array}\right)\,.
\end{equation}
The b.c. in Eq.~(\ref{bc}) is covariant only under the $SU(2)\times U(1)$ subgroup of $SU(3)$
which commutes with $P$ and the gauge group is then restricted to the EW one in $4$ dimensions.
By decomposing Eq.~(\ref{bc}) in $SU(2)\times U(1)$-covariant and $4$ dimensional notation we find 
that the ${\bf Z_N}$ twists of each field are
\begin{eqnarray}
&&A_{\mu,\, a}:1\;,\;A_{\mu\,\pm i}:\tau^{\pm1}\;,\;A_{z,a}:\tau^{-1}
\;,\;A_{z,\pm i}:\tau^{-1\pm1}\,,\nonumber \\
 &&{a=\{1,2,3,8\}\,,\;\;\; \pm1=4\mp i5\,,\;\pm2=6\mp i7}\,, 
\end{eqnarray}
where $a\in {\bf 3_0}\oplus {\bf 1_0}\,,\;\pm i\in \bf 2_{\pm \frac12}$. We see that, for 
$N\neq 2$,  
just one doublet of scalar zero-modes is left by the projection. From now on, we will 
focus on this case \footnote{A similar set-up, in the case $N=2\,$, has been considered 
in \cite{ABQ}.} and identify this zero-mode as the Higgs, $A_{z,+i}^0\equiv H_i$, while the 
EW bosons are provided by the zero modes of $A_{\mu,a}$.

\subsection{The tree-level Higgs potential}

The classical Lagrangian for the zero modes contains a quartic coupling for the Higgs
$$
-\frac12{\rm Tr}F^2\;\longrightarrow\; 
-g^2{\rm Tr} [A_{z}^0,A_{\bar z}^0]^2=\lambda(H^\dagger H)^2\,, 
$$
whose strength is related by gauge invariance to the gauge coupling $g$
 $$
\lambda=\frac{g^2}2\,.
$$ 
Of course, an Higgs mass-term 
cannot be present in the classical action, but it will be induced by radiative corrections. To study
the EWSB, say, at one loop, we can consider a potential
\begin{equation}
V(H) = - \mu^2 |H|^2 + \lambda |H|^4\,,
\label{pot}
\end{equation}
having neglected all higher powers of $H$ and assumed EWSB to occur, i.e. the 1-loop Higgs mass-term 
to be negative. Notice that in Eq.~(\ref{pot}) we give the quartic coupling constant its tree-level 
value being the one loop correction small.

From this potential we get masses for the Higgs boson and the $W$
\begin{equation}
m_H = \sqrt{2}v\sqrt{\lambda}\,,\;\;\;\;\;m_W=1/2 g v\,,
\label{mas}
\end{equation}
where $v=\mu/\sqrt{\lambda}=\sqrt{2}\langle|H|\rangle$. If we make use of our tree-level relation 
among $\lambda$ and $g$, we find the Higgs-W mass ratio to be
\begin{equation}
\frac{m_H}{m_W} = 2\,.
\label{mratio}
\end{equation}
We then see how this mass ratio is increased in single-Higgs \footnote{In the $2$ Higgs case the 
situation is different. Due to the presence of flat directions in the classical potential, 
we will in general find states with smaller masses.} 
$6D$ models with respect to the $5D$ 
ones. This happens as, look at Eq.~(\ref{mas}), $\lambda$ would be 1-loop generated in $5D$ models,
while it is tree-level in $6D$. Therefore, it is bigger.

\subsection{The tadpole operator}
Unfortunately, the increase of the Higgs mass is not the only new feature of $6$ dimensional models.
 At the fixed points of the orbifold the theory is not invariant under all the $SU(3)$ local transformations, 
but just
under those belonging to the $SU(2)\times U(1)$ subgroup. Such transformations act trivially on the $U(1)$ 
generator, which corresponds to the ``$8$'' direction in the $SU(3)$ embedding.
Therefore, the localized ``tadpole'' operator
\begin{equation}
F_{z\bar{z}}^8\,\delta(z-z_f)=\left[\partial_zA_{\bar z}^8-\partial_{\bar z}A_{z}^8+g_6f^{8bc}
A_{z\,b}A_{\bar{z}\, c}\right]\,\delta(z-z_f)\,,
\label{tad}
\end{equation}
is allowed by gauge invariance. One can see that it is quadratically divergent at one loop and it 
gives rise to a mass-term for the Higgs field in 4 dimensions.
One can check by direct computation that this operator is indeed generated already in the pure gauge 
theory.

If the tadpole operator is generated at one loop, its contribution to the Higgs mass and to the EWSB 
scale, as estimated by NDA, is too big and the little hierarchy problem is reintroduced. On the contrary, 
if it arises at two loops, its contribution would be small enough not to destabilize the EWSB scale. 
Therefore, we would like to achieve an accidental one loop cancellation by adding a suitable bulk 
fermion content. We have computed the contribution to the tadpole from fermions and scalars in a 
generic representation of the gauge group. We find that the tadpole cannot be canceled at one loop 
without introducing fundamental scalars.

However, we still have the possibility of getting a stable hierarchy from $6D$ models. Let us consider in 
more detail the structure of the tadpole. It is distributed on all the various fixed points of the 
$T^2/{\bf Z_N}$ orbifold
\be
{\mathcal L}_{\rm tad}=\sum_{\rm fix-points }{ \mathcal C}_iF_{z\bar{z}}^8\,\delta(z-z_i)\,,
\label{Ltad}
\ee
and for $N=3,4,6$ there are respectively $1,2,3$ independent coefficients. This is due to discrete 
space-time symmetries of the $T^2/{\bf Z_N}$ orbifolds. 
If we look at Eq.~(\ref{tad}) and Eq.~(\ref{Ltad}) and we remember that the Higgs field has a constant 
wave function along the extra dimension, we see that the Higgs mass-term generated by the tadpole will 
be proportional to the ``integrated'' tadpole coefficient 
${\mathcal C}\equiv \sum_i {\mathcal C}_i$. From our explicit computation of the tadpole we find 
that we cannot make each of the ${\mathcal C}_i$ vanishing at one loop, but we can cancel 
${\mathcal C}$  by a suitable choice of the fermion content, in the $N=4$ case.

The absence of a direct contribution to the Higgs mass-term from a globally vanishing tadpole 
is just a hint that in this case the EWSB may not be destabilized. From Eq.~(\ref{tad}) we see 
that the tadpole also generates a complicate mass matrix for the fields in the Kaluza-Klein tower
of the Higgs (also mixings \emph{with} the Higgs are present) and a tadpole term for $A_{z}^8$ that will 
induce a non trivial background for it. Therefore, in order to see 
if and how much the EWSB is sensible to the divergence in the tadpole coefficients, one has 
 to compute the background induced by the tadpole and then study the quantum fluctuations around it.

\subsection{Globally vanishing tadpole and EWSB}
Let us add the tadpole operator to the tree-level Lagrangian
\be
{\mathcal L}=-\frac12 {\rm Tr}F^2+{\mathcal L}_{\rm tad}\, ,
\ee
and look at the background which minimizes the corresponding action. We need field configurations that 
are invariant under the $4$ dimensional Lorentz group, so we put to zero all $4D$ vectors and just 
retain the $z\,,\; \bar z$ dependence of the scalars. After fixing the $6D$ gauge invariance
we get a potential
\be
V = \frac 12 \sum_{a=1}^3|F_{z 
\bar z}^a|^2 + |F_{z \bar z}^{-i}|^2+ \frac 12 \Big|F_{z \bar z}^8 - 
i\sum_{i}{\mathcal C}_i \delta^{(2)}(z-z_{i})\Big|^2 \,,
\ee
which is a sum of squares, as it happens in supersymmetric theories.
An absolute minimum is found by imposing each term to vanish, and solving then simple first order 
equations. A solution is obtained by putting to zero all the fields that are charged under 
$SU(2)\times U(1)$
\be
 \langle A_{z}^{1,2,3}\rangle=0\,,\;\;\;\;\;\langle A_{z}^{\pm i}\rangle=0\,,
\ee
 if $\langle A_{z}^8(z)\rangle$, which is neutral, satisfies the equation
\be
{F_{z \bar z}^{8}}_{|_{A_{z}^{\pm i}=0}} 
=({d\langle A^8\rangle })_{z\,\bar{z}}=i\sum_{i}{\mathcal C}_i \delta^{(2)}(z-z_{i})\,.
\label{eqaz}
\ee
By integrating both sides of Eq.~(\ref{eqaz}) and using the Stokes' theorem we immediately see 
that it cannot have a solution if the tadpole is not globally vanishing. 
Defining $A_{z}^8\equiv i\partial_{\bar z}W$ we find for $W$ a Poisson equation whose solution is 
a linear combination of scalar Green functions on the internal torus. The explicit form of this solution 
can be found in \cite{zuck}.

As the background induced by a globally vanishing tadpole is \emph{neutral} under the EW group, we see 
that \emph{a globally vanishing tadpole does not induce EWSB}. 

Consider now quantum fluctuations around this background. We find that the zero-mode wave function for 
the field $A_{z}^{-i}$ satisfies the equation
\begin{equation}
F_{\bar z z}^{-i}= \bigg(\partial_{\bar 
z}+ i g_6 \tan \theta_W \frac 12 \langle A_z^{8}\rangle \bigg) 
A_{z,0}^{-i} = 0\,,
\end{equation}
which is easily seen to have solution 
\begin{equation}
A_{z,0}^{-i} = \displaystyle{e^{g_6 \tan \theta_W \frac 12 W }}\,.
\label{Hwf}
\end{equation}
Therefore, we see that \emph{a globally vanishing tadpole does not contribute to the Higgs mass}.

\section{Conclusions}

We investigated Gauge-Higgs Unification models in $6$ Dimensions based on the $SU(3)$ gauge group.

We have seen that by $T^2/{\bf Z_N}$ compactification (for ${\bf N}\neq 2$) single-Higgs models can be 
obtained. In this case, the tree-level quartic coupling which is induced for it gives an important 
effect in increasing the Higgs mass with respect to the W-boson one.

The tadpole operator which may give rise to the destabilization of the EWSB scale has been identified and 
computed at one loop. The result is that it cannot be completely canceled by adding bulk fermions but, 
in the $T^2/{\bf Z_4}$
case, it can be made globally vanishing.

We have shown that in the globally vanishing case the presence of the divergence in the tadpole 
coefficients could be tolerated in a realistic model. The tadpole coefficients, indeed, does not 
contribute either to the EWSB scale or to the Higgs mass that would be radiatively induced by 
 non-local operators, as if the tadpole was absent. The difference is of course that now the 
wave function of the Higgs has a non trivial profile in the extra dimensions (see Eq.~(\ref{Hwf}))
and that the theory lives in a complicate $A_{z}^{8}$ background which would make computations 
harder.

Anyway, there is still an obstruction in building realistic models in this set-up. 
The 1-loop cancellation of the integrated tadpole poses a restriction on the bulk fermion content 
\footnote{Localized fermions, as the SM ones, does not contribute to the quadratic divergence of the 
tadpole, so we have no constrains on them.} 
and another restriction comes from the condition that the extra-dimensional gauge group should be 
anomaly free, or a least it should be possible to cancel the anomaly by some Green-Schwarz mechanism.
It may be difficult to find a spectrum satisfying both conditions.

\end{document}